\documentclass{iopart}
\usepackage{iopams}
\usepackage{cite}

\newcommand{\bear}{\begin{eqnarray}}
\newcommand{\eear}{\end{eqnarray}}
\newcommand{\be}{\begin{equation}}
\newcommand{\ee}{\end{equation}}
\newcommand{\beqn}{\begin{eqnarray}}
\newcommand{\eeqn}{\end{eqnarray}}
\newcommand{\beqnn}{\begin{eqnarray*}}
\newcommand{\eeqnn}{\end{eqnarray*}}

\def\vep{\varepsilon}
\def\vf{\varphi}

\begin{document}

\title[Relative energy difference of Gaussian states with a fixed fidelity]
{Upper bounds on the relative energy difference of 
pure and mixed Gaussian states with a fixed fidelity}

\author{V V Dodonov }


\address{
 Instituto de F\'{\i}sica, Universidade de Bras\'{\i}lia,
Caixa Postal 04455, 70910-900 Bras\'{\i}lia, DF, Brazil }

\begin{abstract}

Exact and approximate formulas for the upper bound of the relative energy difference of two 
Gaussian  states with the fixed fidelity between them are derived. The reciprocal formulas for
the upper bound of the fidelity for the fixed value of the relative energy difference are
obtained as well. 
The bounds appear higher for pure states than for mixed ones, and their maximal values
correspond to squeezed vacuum states. In particular, to guarantee
the relative energy difference less than 10\%, for quite arbitrary Gaussian states, 
the fidelity between them must exceed the level 0.998866. 
\end{abstract}

\ead{vdodonov@fis.unb.br} 

\submitto{\JPA}

\section
{\em Introduction}

There are many areas of quantum physics 
(e.g., quantum teleportation or quantum state engineering)
where one has to compare somehow different quantum states.
In particular, in many cases it is important to know, how `close' are two quantum states
(e.g., the initial one and its teleported or reconstructed partner),
described by the statistical operators $\hat\rho_1$ and $\hat\rho_2$.
An accepted measure of closeness 
is the Bures--Uhlmann fidelity \cite{Joz}
\be
{\cal F} =\left[\mbox{Tr}\left(\sqrt{\sqrt{\hat\rho_1} \hat\rho_2 \sqrt{\hat\rho_1}}\right)\right]^2.
\label{fid}
\ee
Nowadays one can meet this quantity in almost every paper on quantum information, and many
interesting results have been obtained for the past decade.
For example, various boundaries or critical values for fidelities, corresponding to different 
teleportation protocols of some specific classes of quantum states, were derived in 
\cite{Braun00,Cerf00,Gran01}. In particular, the value ${\cal F}=1/2$
has been established as the boundary between classical and quantum domains in the teleportation of
coherent states of the electromagnetic field \cite{Braun00}, while another critical value ${\cal F}=2/3$ was found in
\cite{Cerf00,Gran01}. The meaning of these numbers was further elucidated in \cite{Ban04,Cav04}.
Experimentalists also frequently use the fidelity as measure of quality of their achievements.
A remarkable progress is observed for the past decade: the reported experimental values increased from
$0.58$ \cite{Furu98}, $0.62$ \cite{Zhang03} and $0.64$ \cite{Bow03} to 
$0.87$ \cite{Reed10}, $0.93$ \cite{Specht11} and $0.98$ \cite{Zavat09}.

However, looking at these numbers, the following inevitable question arises: 
is it sufficient to have, say, $90$\%  fidelity (or even $98$\%), to be sure that the two states are indeed 
`similar' or `close' to each other?
It seems that the answer depends on the concrete situation and additional information or assumptions about the states,
in particular, on the exact meaning of the word `close'. 

The essence of the problem and motivation to study it can be elucidated in the following simple example.
Consider two coherent states $|\alpha\rangle $ and $|\beta\rangle $. Their fidelity equals
${\cal F}=|\langle\alpha|\beta\rangle|^2=\exp\left(-|\alpha-\beta|^2\right)$, so it depends on the difference
$|\alpha-\beta|$ only. 
The question is: are the two pairs of states, $\left\{|\alpha\rangle , |\beta\rangle \right\}$ and
$\left\{|\alpha +A\rangle , |\beta +A\rangle \right\}$, `equally close' (or `equally distant')?
From the pure geometrical point of view, the answer is definitely positive, since the second pair can be
obtained from the first one by means of a unitary transformation or by a simple shift in the complex plane
of parameters $\alpha$. From this point of view, the space of parameters is homogeneous.
But it is certainly non-homogeneous from the physical point of view, since the vacuum state $|0\rangle$
is obviously distinguished (in this connection, the `energy-sensitive' distances between quantum
states were proposed in \cite{dod99}). 
Let us take $|\alpha-\beta|=1$. Then ${\cal F}=0.37$, which certainly seems to be 
a low value. Accordingly, the states with $\alpha=0$ and $\beta=1$ seem very different. But what
one can say about the
states with $\alpha=1000$ and $\beta=1001$? Their fidelity is low, but in many (of course, not all) cases
replacing one of these two states with another hardly would cause a significant change in experimental results
or their interpretation. 
On the other hand, the value of ${\cal F}=0.9$ corresponds to the difference $|\alpha-\beta|=0.325$, and
it is not so obvious now, whether the coherent state with $\alpha=0.325$ can be considered as a good
approximation of the vacuum state? These examples show that the fidelity alone not always can be used
as an adequate quantity characterizing the degree of `similarity' (`closeness') of quantum states. In many cases 
some other quantities should be taken into account in addition to the fidelity.

One of the most important physical parameters is the energy. Therefore it seems interesting to answer the
following question: how large the energy difference between two quantum states with the given fidelity can be?
If it is small and the fidelity is high, then one may have more firm reasons to say that two states are `close'.
On the contrary, if the energy difference turns out to be big, then hardly two states can be considered as
`close', even if the fidelity is rather high.

One can easily see that there is no unique answer to the question put above in the most general case. 
Indeed, let us consider two states,
$|\psi\rangle$ and $|\chi\rangle =\sqrt{F}|\psi\rangle +\sqrt{1-F}|\psi_{\perp}\rangle$, where 
the state $|\psi_{\perp}\rangle$ is orthogonal to $|\psi\rangle$. The fidelity between these states
equals ${\cal F}=|\langle\psi|\chi\rangle|^2 = F$. Since the space of states $|\psi_{\perp}\rangle$
is very big (I have in mind the continuous variable systems, when the Hilbert space of quantum states is
infinite-dimensional), one can expect that there exist such states $|\psi_{\perp}\rangle$ that result
in arbitrarily large difference of mean energies in the states $|\psi\rangle$ and $|\chi\rangle$,
without changing their fidelity (this is obvious if $|\psi\rangle$ is the Fock state).

But if one restricts somehow the space of admissible quantum states, then the above reasonings may become
invalid (for example, there are no orthogonal coherent or squeezed states), and some universal (for the
selected family) bounds can be found. It appears, in particular, that this is just the case for the family of 
{\em Gaussian\/}  states. Since these states are frequently used in the contemporary 
theoretical and experimental quantum optics,
and since the final results turn out remarkably simple, I believe that these results could be interesting
for many readers.

\section
{\em Pure Gaussian states}

It is convenient to start from the simplest case of two {\em pure\/}
squeezed quantum states 
described by the normalized  wave functions
\[
\psi_j(x) = \left(a_j/\pi\right)^{1/4} \exp\left[
-\frac12\left(a_j +ib_j\right)\left(x-x_j\right)^2 + ip_j x\right],
\]
where $a_j>0$, whereas real parameters  $b_j$, $x_j$ and $p_j$ can assume arbitrary values ($j=1,2$).
 For pure states, the fidelity (\ref{fid}) is
 reduced to the scalar product $
{\cal F}=\left|\langle\psi_1|\psi_2\rangle\right|^2$,
which 
can be easily calculated:
\be
{\cal F} = 2\sqrt{a_1 a_2/G}\exp\left(- U/G\right),
\label{Fsqz}
\ee
\be
G= \left(a_1 +a_2\right)^2 + \left(b_1 -b_2\right)^2,
\label{defG}
\ee
\be
\fl
U = \left(a_1 +a_2\right)(\delta p)^2 
+ 2\left(a_1 b_2 + a_2 b_1\right)(\delta p)(\delta x) 
+ 
\left[a_1 a_2 \left(a_1 +a_2\right) +a_1 b_2^2 + a_2 b_1^2\right](\delta x)^2,
\label{defU}
\ee
where $\delta x = x_2 - x_1$ and $\delta p = p_2 - p_1$.

I suppose that the states $\psi_j(x)$ describe a quantum oscillator with unit mass and frequency.
Then the mean energy of each state equals (assuming $\hbar=1$)
\be
 E_j = \left(p_j^2 + x_j^2\right)/2 + \left(1+a_j^2 +b_j^2\right)/\left(4a_j\right).
 \label{Ej}
\ee
The question is: how big can the difference 
$ \delta E = E_2 -  E_1$ be 
for the two states with different parameters but the fixed fidelity ${\cal F}$?
To simplify the formulas, it is convenient to introduce the following notation:
$a_1\equiv a$,
$ a_2 -a_1 = a\alpha$, $b_1=ac$, $b_2 -b_1 = a\beta$.
Then
\be
\delta E = p(\delta p) +x(\delta x) + \frac12\left[(\delta p)^2 + (\delta x)^2\right]
+B,
\label{delE}
\ee
\be
B(a,c,\alpha,\beta) = \frac{a^2\left[\alpha(1+\alpha -c^2) +\beta(2c +\beta)\right] -\alpha}
{4a(1+\alpha)}.
\ee

Although the parameters $\alpha$, $\beta$, $\delta x$ and $\delta p$ are limited
 for the fixed fidelity ${\cal F}$, the energy difference $\delta E$ can assume arbitrarily
big values, if the initial values $x$ and $p$ are very big. The same can happen if parameter 
$a$ is very big or very small. Therefore it seems reasonable to study the limits of
variations of the {\em relative change\/} of energy
\be
{\cal E} = \delta E/ E_1 = E_2/E_1 -1.
\label{defcalE}
\ee
For the fixed values of $\delta p$, $\delta x$ and $ E_1$,
the sum $p(\delta p) +x(\delta x)$ is maximal if the vectors $(\delta x, \delta p)$ and
$(x, p)$ are parallel, and it is minimal for anti-parallel vectors. 
Therefore we have to find the maximal (positive) and minimal (negative) values of the functions
\be
{\cal E}_{\pm} = \frac{\delta^2/2 \pm R\delta +  B}
{R^2/2 + A}, 
\label{E-RAB}
\ee
where
$
A(a,c) = \left[1+a^2\left(1+c^2\right)\right]/(4a)$,
$ R =  \sqrt{x^2 +p^2}$, $\delta = \sqrt{(\delta x)^2 +(\delta p)^2}$.
Calculating the extremal values of (\ref{E-RAB}) with respect to parameter $R$, 
 we arrive at the problem of finding extremal values of functions
\be
{\cal E}_{\pm} = \pm\,\frac14\left[\sqrt{\left(\Delta +2L\right)^2 +8\Delta} \pm \left( \Delta +2L\right)\right]
\ee
under the constraint 
\[
 \sqrt{\frac{4(1+\alpha)}{4(1+\alpha)+\alpha^2+\beta^2}}\,
\exp\left[-\, \frac{ \tilde{u}(\vf,a,c,\alpha,\beta)}{g(a,\alpha,\beta)}\,\delta^2
\right]
= {\cal F}.
\]
Here $\Delta = \delta^2/A$, $L=B/A$,
$g= a\left[(2+\alpha)^2+\beta^2 \right]$, 
\beqnn
\tilde{u} &=&
(2+\alpha)\sin^2(\vf) +
a^2\left[(1+\alpha)(2+\alpha +c^2) 
+(c +\beta)^2 \right]\cos^2(\vf)
 \\&& 
+ 2a\left(2c +\beta +c\alpha\right) \sin(\vf)\cos(\vf)
\eeqnn
and the new free parameter $\vf$ is the angle between the direction of vector $(\delta x, \delta p)$
and the horizontal axis.

\subsection{The case of different displacements but identical squeezing parameters}

There are two special cases. 
The first one is $\alpha=\beta=0$ (two squeezed states with different displacement parameters
but identical variances of quadratures). Then 
\be
{\cal E}_{\pm} = \pm\,\frac14\left(\sqrt{\Delta^2 +8\Delta} \pm \Delta \right),
\label{E-del}
\ee
\[
\Delta = \frac{8a^2 f \left[
1+a^2\left(1+c^2\right)\right]^{-1}}{\sin^2(\vf) +
a^2\left(1+c^2\right)\cos^2(\vf) +2ac \sin(\vf)\cos(\vf)},
\]
where
$f =\ln(1/{\cal F})$.
The maximal positive value of function (\ref{E-del}) and its minimal negative value
(in the case of sign ``minus'') are achieved for the maximal possible value of the positive
coefficient $\Delta$.
Looking for extrema of $\Delta$ as function of $\vf$ 
one obtains
$
\Delta_{max-\vf} = 4f\kappa$,
where 
\be \kappa= 1 + \sqrt{1-\xi}, \qquad
\xi(a,c)={4a^2}/
{\left[a^2\left(1+c^2\right) +1\right]^2}.
\ee
Then equation (\ref{E-del}) results in the following limitations on the relative energy difference
of two states with fixed parameters $a$ and $c$ (but arbitrary displacement parameters)
for the given value of fidelity:
\be
-\,\frac{2\sqrt{f\kappa}}{\sqrt{f\kappa+2} + \sqrt{f\kappa}} \le {\cal E} \le 
\frac{2\sqrt{f\kappa}}{\sqrt{f\kappa+2} - \sqrt{f\kappa}}.
\label{E-kap}
\ee
One can notice an asymmetry between the left-hand and right-hand sides of inequality (\ref{E-kap}).
Its origin is in the definition (\ref{defcalE}), which is asymmetrical with respect to
the states $|\psi_1\rangle$ and $|\psi_2\rangle$.
Due to this definition one has the
restriction ${\cal E} >-1$ for negative values of ${\cal E}$ (since the total energy of the state $|\psi_2\rangle$ is positive),
whereas there are no bounds for positive values of ${\cal E}$.
The asymmetry in (\ref{E-kap}) disappears if one introduces the {\em symmetric relative energy difference\/}
\be
{\cal Y} = \frac{\left|E_2 -E_1 \right|}{\sqrt{E_1 E_2}} =\frac{|{\cal E}|}{\sqrt{1+{\cal E}}}.
\label{defY}
\ee
Then both the inequalities in (\ref{E-kap}) lead to the inequality
\be
{\cal Y} \le \sqrt{2f\kappa}.
\label{Y-kap}
\ee
One could normalize the difference $\left|E_2 -E_1 \right|$ in (\ref{defY}) not by the `geometrical' mean value
$\sqrt{E_1 E_2}$, but, say, by the usual `arithmetical' mean value $\left(E_1+ E_2\right)/2$, but the resulting
formulas are much more complicated and less attractive than the simple inequality (\ref{Y-kap}).
It would be interesting to know, whether there are some deep physical or geometrical reasons for choosing the geometrical mean values
instead of arithmetical ones, or this is some mathematical artefact.

In the special case of $a=1$ and $c=0$ (when $\kappa=1$) we obtain the following
{\em exact\/}  relation between the {\em maximal\/}
symmetric relative energy difference
${\cal Y}_{m}$
and the fidelity between two {\em coherent states}:
\be
{\cal Y}_{m}^{(coh)} = \sqrt{2\ln(1/{\cal F})}.
\label{YFcoh}
\ee
The inverse relation gives the {\em maximal fidelity\/} between two coherent states
for the given value of ${\cal Y}$:
\be
{\cal F}^{(coh)}_{max} = \exp\left(-{\cal Y}^2/2\right).
\label{FYcoh}
\ee

For {\em arbitrary\/} squeezed states with equal squeezing coefficients $a$ and $c$, but different
displacement parameters, one should take into account that 
coefficient $\xi$ goes to zero (and $\kappa \to 2$) for highly squeezed states: either if $a\gg 1$
(strong coordinate squeezing), or $a\ll 1$ (strong momentum squeezing), or $|c|\to \infty$ for a fixed $a$
(strongly correlated states). 
For this family of states (labeled by the supescript $\delta$) we obtain the relations
\be
{\cal Y}_{m}^{(\delta)} = 2\sqrt{\ln(1/{\cal F})},
\qquad
{\cal F}^{(\delta)}_{max} = \exp\left(-{\cal Y}^2/4\right). 
\label{YFdel}
\ee

\subsection{Undisplaced squeezed states}

Another special case is $\delta =0$.
Then we have  the constraint
\be
\left(\alpha^2 +\beta^2\right)/(1+\alpha) = 4 D, \quad
 D =\left(1-{\cal F}^2\right)/{\cal F}^2.
 \label{D}
 \ee
Formula (\ref{E-RAB}) shows that the maximal relative energy difference ${\cal E}$ in this case can be achieved
 for {\em undisplaced squeezed states} with $R=0$. It can be written as
\be
{\cal E}= {\cal E}_{\alpha} + 2\chi(K + Mc),
\label{EKM}
\ee
where
$\chi = {a^2}/\left[1 +a^2\left(1+c^2\right)\right]$
does not depend on  $\alpha$ and $\beta$, whereas
 ${\cal E}_{\alpha}$, $K$ and $M$ do not depend on  $a$ and $c$: 
\[
{\cal E}_{\alpha}=-\,\frac{\alpha}{1+\alpha}, \quad
K= 2D -{\cal E}_{\alpha}, \quad
M= \frac{\beta}{1+\alpha}.
\]
For a fixed parameter $c$, the coefficient $\chi$ varies from $0$ to 
$\left(1+c^2\right)^{-1}$. Consequently, ${\cal E}$ can vary between 
${\cal E}_{\alpha}$ and 
$
{\cal E}_c = {\cal E}_{\alpha} + {2(K+Mc)}/\left(1 +c^2\right)$.
The extremal points of ${\cal E}_c$ as function of variable $c$ 
are given by the roots of the  equation
$Mc^2 + 2Kc -M=0$, i.e.,
$
M c_{\pm} = -K  \pm \sqrt{K^2 +M^2}$.
These roots give us immediately  the following
extremal values of function ${\cal E}_c(c)$:
\[ 
{\cal E}_{c\pm} = {\cal E}_{\alpha} \pm\left(\sqrt{K^2 +M^2} \pm K\right)
= 2D \pm 2\sqrt{D(1+D)}.
\]
Surprisingly, these values do not depend on $\alpha$ or $\beta$.
They should be compared with the values of ${\cal E}_{\alpha}$ corresponding to
 maximal and minimal possible values of $\alpha$ for the fixed $\beta$, namely
$
\alpha_{\pm}(\beta) = 2D \pm \sqrt{4D(1+D) -\beta^2}$.
Obviously, the minimal (negative) value of $\alpha_{-}(\beta)$ (achieved for $\beta=0$)
gives the maximal (positive) value of ${\cal E}_{\alpha}$, whereas the maximal (positive)
value of $\alpha_{+}(\beta)$ (also achieved for $\beta=0$) yields 
the minimal (negative) value ${\cal E}_{\alpha}$. One can verify that these two extremal values
coincide exactly with ${\cal E}_{c\pm}$.
Thus we arrive at the inequalities
\be
-\,\frac{2\sqrt{1-{\cal F}^2}}{\sqrt{1-{\cal F}^2} + 1} <
{\cal E} < \frac{2\sqrt{1-{\cal F}^2}}{1 -\sqrt{1-{\cal F}^2} },
\label{E-F}
\ee
which are equivalent to the following relations between the maximal symmetric relative energy difference
${\cal Y}_m$ (\ref{defY})
for the given fidelity ${\cal F}$ and the maximal possible fidelity ${\cal F}_{max}$ 
for the given value of ${\cal Y}$:
\be
{\cal Y}_m = \frac{2\sqrt{1-{\cal F}^2}}{{\cal F}}, \qquad
{\cal F}_{max} = \frac1{\sqrt{1+{\cal Y}^2/4}}
= \frac{\sqrt{1+|{\cal E}|}}{1+|{\cal E}|/2}.
\label{YFmm}
\ee
In the limit of ${\cal Y}_m \ll1$
or $1-{\cal F} \ll 1$ we can write
\be
{\cal Y}_m \approx  \sqrt{8(1-{\cal F})}, \qquad
{\cal F}_{max} \approx 1 -{\cal Y}^2/8.
\label{F-Esmall}
\ee
It is easy to see that
\be
{\cal Y}_{m}^{(coh)} < {\cal Y}_{m}^{(\delta)} < {\cal Y}_{m},
\quad
{\cal F}_{max}^{(coh)} < {\cal F}_{max}^{(\delta)} < {\cal F}_{max}.
\label{chain}
\ee

My conjecture is that the upper bounds (\ref{YFmm}) hold, as a matter of fact, in the most general case, when
all three variations ($\alpha$, $\beta$ and $\delta$) can be different from zero.
Many numerical tests made for different sets of parameters \cite{DH} confirm this conjecture: 
in all the cases the fidelity calculated by formula (\ref{Fsqz})
appeared smaller than the maximal value given by (\ref{YFmm}) with the parameter ${\cal Y}$
calculated by means of equations (\ref{Ej}) and (\ref{defY}). 
But I did not succeed to find an analytical proof, except for the simplest (but, perhaps, the most
important) special case when the difference $1-{\cal F}=\vep$ is small. 
Assuming that all variations ($\alpha$, $\beta$, $\delta x$ and $\delta p$) are small and taking into
account only the leading terms with respect to these variables, one can 
replace the exact formula for the relative energy difference ${\cal E}$ by 
\[
{\cal E} = \frac1{E_1} \left\{ p\delta p +x\delta x + 
\frac{a}{4}\left[\alpha\left(1 -c^2\right) +2c\beta\right]
-\frac{\alpha}{4a}
\right\}
\]
and to maximize this function under the constraint
\[
\alpha^2 +\beta^2 +\frac4{a}
\left[ \delta p^2 +
a^2\left(1 +c^2\right) \delta x^2
+ 2ac \delta p\delta x\right]
=8\vep .
\]
This problem can be solved with the aid of the Lagrange multiplier,
and the result coincides with (\ref{F-Esmall}).

\section
{\em Mixed Gaussian states}

Since the extremal values of the fidelity and relative energy difference for pure Gaussian states are observed for 
zero displacement parameters [equation (\ref{chain})], it seems reasonable to suppose that the same is true for mixed
Gaussian states, as well. Therefore let us consider mixed {\em homogeneous\/} Gaussian states described by the density
matrices (kernels of the statistical operators $\hat\rho_k$ in the coordinate representation)
\be
\fl
\rho_k(x,y)= \sqrt{\frac{a_k-z_k}{\pi}}
\exp\left[ -\frac12\left(a_k+ib_k\right)x^2 
-\frac12\left(a_k-ib_k\right)y^2 +a_k\zeta_k xy\right], 
\label{rhok}
\ee
where $a_k$, $b_k$ and $\zeta_k$ are real numbers obeying the inequalities $a_k >0$ and $0\le \zeta_k <1$.
The states (\ref{rhok}) are normalized as follows:
$
\mbox{Tr}\left(\hat\rho_k\right) \equiv \int \rho_k(x,x)\,dx =1$.
The parameters $\zeta_k$ are responsible for the `quantum purity':
\be
\mu_k \equiv \mbox{Tr}\left(\hat\rho_k^2\right) 
= \sqrt{\frac{1-\zeta_k}{1+\zeta_k}}, \qquad \zeta_k =\frac{1-\mu_k^2}{1+\mu_k^2}.
\label{purity}
\ee
The mean energy in the state (\ref{rhok}) equals
\be
 E_j = \frac{1+a_j^2\left(1-\zeta_j^2\right) +b_j^2 }{4a_j\left(1 -\zeta_j\right)}.
\ee
The calculation of the fidelity (\ref{fid}) between mixed Gaussian states is reduced to a chain of Gaussian integrals,
since the square root of the Gaussian positive definite operator is also a Gaussian positive definite operator,
whose kernel can be found using the scheme exposed, e.g., in \cite{Scu}. 
Following this scheme one can arrive after some tedius calculations to the following 
generalization of formula (\ref{Fsqz}) for homogeneous mixed Gaussian states:
\be
{\cal F} =\frac{2\sqrt{a_1 a_2 \left(1-\zeta_1\right)\left(1-\zeta_2\right)}}
{\sqrt{G -\left(a_1\zeta_1 -a_2\zeta_2\right)^2}
-2\sqrt{a_1 a_2 \zeta_1 \zeta_2}}\,,
\ee
where the coefficient $G$ is given by equation (\ref{defG}).
Equivalent formulas for different parametrizations of the Gaussian states were found, e.g., in 
\cite{Ban04,Scu,Twam,MarMarScu,Nha}.

\subsection{Two states with identical fixed purities}

Let us consider first the case of two states having the same fixed value of purity: 
$\zeta_1=\zeta_2=\zeta=const$. Then one has to find the extremal values of the relative
energy difference ${\cal E}(\alpha,\beta)$ under the constraint
\be
\frac{\alpha^2\left(1-\zeta^2\right) +\beta^2}{1+\alpha} = 4 D,
\qquad
D =\frac{(1-{\cal F})}{{\cal F}^2}
(1-\zeta)\left[1-\zeta +{\cal F}(1+\zeta)\right].
\ee
The function ${\cal E}$ in this case has the same form as in (\ref{EKM}), but with 
\[
K= 2D +\frac{\alpha\left(1-\zeta^2\right)}{1+\alpha}, \qquad
\chi = \frac{a^2}{1 +a^2\left(1 -\zeta^2 +c^2\right)}
\]
and unchanged coefficients ${\cal E}_{\alpha}$ and $M$.
 Using the same scheme as before,
one can obtain the formula
\be
{\cal Y}_m({\cal F},\zeta) = \frac{2}{{\cal F}}\sqrt{\frac{1-{\cal F}}{1+\zeta}\left[1+{\cal F} -\zeta(1-{\cal F})\right]}.
\label{Y-mix}
\ee
Obviously, the right-hand side of (\ref{Y-mix}) decreases monotonously when $\zeta$ increases from
$0$ to $1$. Consequently, the most strong bounds on ${\cal Y}_m$ and ${\cal F}_{max}$ take place for
{\em pure\/} quantum states ($\zeta=0$), and they are given again by equation (\ref{YFmm}).
In particular, for $\zeta \to 1$ (`supermixed' states with the purity $\mu \to 0$) we have
\be
{\cal Y}_m^{smix}({\cal F}) = {2}\sqrt{\frac{1-{\cal F}}{{\cal F}}}, 
\quad 
{\cal F}_{max}^{smix}({\cal Y}) = \frac{1}{1+{\cal Y}^2/4}.
\label{F-supermix}
\ee
For small values of $1-{\cal F}$ and ${\cal Y}$
\be
{\cal Y}_m^{smix} \approx  2\sqrt{1-{\cal F}}, \qquad
{\cal F}_{max}^{smix} \approx 1 -{\cal Y}^2/4.
\label{F-Esmall-supermix}
\ee

\subsection{Deviations from the pure state}

Another case which can be treated analytically is $\zeta_1=0$ and $\zeta_2\equiv \zeta \ge 0$.
Then the constraint has the form
\be
\frac{(2+\alpha)^2 +\beta^2 -\zeta^2(1+\alpha)^2}
{(1+\alpha)(1-\zeta)} = \frac4{{\cal F}^2}.
\label{F-al-mix0}
\ee
The function ${\cal E}$ can be written again in the form (\ref{EKM}) with the same coefficient
$\chi$, but  three other coefficients are different:
\[
\fl
{\cal E}_{\alpha}= \frac1{(1+\alpha)(1-\zeta)} -1, \qquad
K = \frac2{{\cal F}^2} -\frac{2+\alpha}{(1+\alpha)(1-\zeta)},  \qquad
M = \frac{\beta}{(1+\alpha)(1-\zeta)}.
\]
Then the same scheme as before leads to the formula 
\[
\fl
{\cal Y}_m = \frac{1}{{\cal F}\tilde\zeta}
\left( \sqrt{2\left(1 - \zeta\right)\left(1 + \tilde\zeta\right) -{\cal F}^2\left(1 + \tilde\zeta\right)^2}
+ \sqrt{2\left(1 - \zeta\right)\left(1 - \tilde\zeta\right) -{\cal F}^2\left(1 - \tilde\zeta\right)^2}
\right)
\]
where $\tilde\zeta =\sqrt{1-\zeta^2}$.
For $\zeta\ll 1$ we have 
\be
{\cal Y}_m \approx \frac{1}{{\cal F}}\left(2\sqrt{1-{\cal F}^2 -\zeta +\frac{\zeta^2}{4}\left(2{\cal F}^2 -1\right)} 
+\zeta\right) .
\label{Ymzeta}
\ee
The right-hand side of (\ref{Ymzeta}) is obviously
smaller than the pure state bound (\ref{YFmm}).
The maximal possible value of $\zeta$ for the fixed ${\cal F}$ follows from equation (\ref{F-al-mix0})
with $\alpha=\beta=0$:
$
\zeta_{max} = 2{\cal F}^{-2}\left(1-\sqrt{1-{\cal F}^2\left(1-{\cal F}^2\right)}\right)$.
If $\vep=1-{\cal F} \ll 1$, then 
$\zeta_{max} =2\vep -5\vep^3/2 +\cdots$. Putting this critical value in (\ref{Ymzeta}) one can obtain the value
${\cal Y}_{m}(\zeta_{max})=  2\vep$, which is much smaller than the pure state boundary $\sqrt{8\vep}$.

\section{Discussion}

The results obtained are illustrated in two tables below.
The first one gives the maximal possible fidelities between two coherent states
[equation (\ref{YFcoh})],
two displaced pure squeezed states with arbitrary fixed variances [equation (\ref{YFdel})],
two `supermixed' homogeneous squeezed states with $\zeta\to 1$ [equation (\ref{F-supermix})]
and two arbitrary (pure or mixed) squeezed states with zero displacements [equation (\ref{YFmm})] for fixed values
of the energy ratio $E_2/E_1$.
The second table gives the values of the maximal symmetrical energy difference ${\cal Y}_m$ and the corresponding maximal possible ratio
$\left(E_2/E_1\right)_{m}$ for the given value of fidelity
between the most general Gaussian states [equation (\ref{YFmm})].
\begin{center}
\begin{tabular}{|l|l|l|l|l|l|}   \hline
$E_2/E_1$ & ${\cal Y}$ & ${\cal F}_{max}^{(coh)}$ & ${\cal F}_{max}^{(\delta)}$ & ${\cal F}_{max}^{smix}$ & ${\cal F}_{max}$ \\  \hline
3         & 1.155      &  0.51                    & 0.72                      & 0.75             & 0.87          \\  \hline
2         & 0.707      &  0.78                    & 0.88                      & 0.89             & 0.94          \\  \hline
1.5       & 0.408      &  0.92                    & 0.96                      & 0.96             & 0.98          \\  \hline
1.1       & 0.095      &  0.9955                  & 0.9977                    & 0.9977           & 0.998866           \\  \hline
\end{tabular}
\begin{tabular}{|l|l|l|}   \hline
${\cal F}$ & ${\cal Y}_m$   & $\left(E_2/E_1\right)_{m}$  \\  \hline
0.999      &  0.09          &  1.09                 \\  \hline
0.99       &  0.28          &  1.32                 \\  \hline
0.95       &  0.66          &  1.91                 \\  \hline
0.9        &  0.97          &  2.55                 \\  \hline
\end{tabular}
\end{center}

It is worth emphasizing that there exist quantum states with the values of ${\cal Y}$ or ${\cal F}$
which are arbitrarily close to the maximal ones found above (for the fixed value of the other parameter).
In particular, ${\cal F}_{max} = \sqrt{8/9} \approx 0.94$ for $ E_2 = 2 E_1$ or $ E_1 = 2 E_2$.
This means that even $94\%$ of fidelity cannot guarantee that two squeezed states are really
close in energy to each other ({\em in the absence of any additional information about the states}). 
Explicit examples are two states with $b_j=x_j=p_j=0$ and $a_2=2a_1$
(then $\alpha=1$) or $a_2=a_1/2$ (then $\alpha=-1/2$), with the fidelity  $94\%$. 
Taking $a_1=1$ (the oscillator ground state),
one obtains the mean energies $\langle E_1\rangle=1/2$ and $\langle E_2\rangle =5/8$,
which are not too different (${\cal Y}=1/\sqrt{20}\approx 0.22$). 
However, for the highly squeezed states with $a_1=0.1$ and
$a_2=0.05$ one has $\langle E_1\rangle = 2.525$ and $\langle E_2 \rangle = 5.0125$; hardly
these two states with ${\cal Y} \approx 0.7$ can be considered as `close'. 
On the other hand, if one knows, for example, that the first state is coherent with $x_1=100$
and $p_1=0$, and the second state is also coherent with $p_2=0$, then ${\cal F}=0.94$ means
that $x_2=100.35$ or $x_2=99,65$. In this case ${\cal Y}\approx 0.007$, which can be accepted as a small quantity,
meaning that these two concrete states are `close'.
But to {\em guarantee\/} that the relative energy difference is below, say, $10\%$ ($E_2\le 1.1 E_1$)
{\em in the absence of any additional information}, 
the fidelity must be higher than 
$\sqrt{440/441} \approx 0.998866$. 

Formulas derived in this paper can be rewritten in terms of the Bures--Uhlmann  distance \cite{Joz}
${\cal B}=\left(2-2\sqrt{{\cal F}}\right)^{1/2}$.
But the corresponding expressions seem less attractive.
 For example, instead of (\ref{YFmm}) one obtains the following formula for the
{\em minimal\/} Bures distance between two Gaussian states with the fixed relative energy difference:
\be
{\cal B}_{min}=\sqrt2\left[1-\left(1+{\cal Y}^2/4\right)^{-1/4}\right]^{1/2}.
\ee

The mutual bounds on the fidelity and the relative energy difference can be derived also for other
interesting families of quantum states. In some cases (e.g., for binomial and negative binomial states)
 analytical formulas can be found \cite{DJRLR}, but in the most of cases (e.g., for superpositions of
coherent states) this can be done only numerically. Perhaps, it would be interesting to obtain
different bounds taking into account other parameters (besides the energy) characterizing quantum states, 
such as the degree of squeezing, Mandel's parameter, or something else.

\section*{Acknowledgment}

The support of the Brazilian funding agency  CNPq is acknowledged.

\end{document}